\documentclass[prb,preprint,showpacs,preprintnumbers,superscriptaddress,amsmath,amssymb]{revtex4}

\usepackage[dvips]{color}

\usepackage{graphicx}% Include figure files
\usepackage{dcolumn}% Align table columns on decimal point
\usepackage{bm}% bold math
\usepackage{latexsym}
\begin{document}

\title{Manifestations of multiple-carrier charge transport in the magnetostructurally ordered phase of undoped BaFe$_2$As$_2$}
% Force line breaks with \\

\author{S. Ishida}
 \affiliation{Department of Physics, University of Tokyo, Tokyo 113-0033, Japan}
 \affiliation{National Institute of Advanced Industrial Science and Technology, Tsukuba 305-8568, Japan}
 \affiliation{JST, Transformative Research-Project on Iron Pnictides, Tokyo 102-0075, Japan}
 
\author{T. Liang}
 \affiliation{Department of Physics, University of Tokyo, Tokyo 113-0033, Japan}
 \affiliation{National Institute of Advanced Industrial Science and Technology, Tsukuba 305-8568, Japan}
 \affiliation{JST, Transformative Research-Project on Iron Pnictides, Tokyo 102-0075, Japan}

\author{M. Nakajima}
 \affiliation{Department of Physics, University of Tokyo, Tokyo 113-0033, Japan}
 \affiliation{National Institute of Advanced Industrial Science and Technology, Tsukuba 305-8568, Japan}
 \affiliation{JST, Transformative Research-Project on Iron Pnictides, Tokyo 102-0075, Japan}
 
\author{K. Kihou}
 \affiliation{National Institute of Advanced Industrial Science and Technology, Tsukuba 305-8568, Japan}
 \affiliation{JST, Transformative Research-Project on Iron Pnictides, Tokyo 102-0075, Japan}
 
\author{C. H. Lee}
 \affiliation{National Institute of Advanced Industrial Science and Technology, Tsukuba 305-8568, Japan}
 \affiliation{JST, Transformative Research-Project on Iron Pnictides, Tokyo 102-0075, Japan}

\author{A. Iyo}
 \affiliation{National Institute of Advanced Industrial Science and Technology, Tsukuba 305-8568, Japan}
 \affiliation{JST, Transformative Research-Project on Iron Pnictides, Tokyo 102-0075, Japan}
 
\author{H. Eisaki}
 \affiliation{National Institute of Advanced Industrial Science and Technology, Tsukuba 305-8568, Japan}
 \affiliation{JST, Transformative Research-Project on Iron Pnictides, Tokyo 102-0075, Japan}
 
\author{T. Kakeshita}
 \affiliation{Department of Physics, University of Tokyo, Tokyo 113-0033, Japan}
 \affiliation{JST, Transformative Research-Project on Iron Pnictides, Tokyo 102-0075, Japan}

\author{T. Kida}
 \affiliation{JST, Transformative Research-Project on Iron Pnictides, Tokyo 102-0075, Japan}
 \affiliation{KYOKUGEN, Osaka University, 1-3 Machikaneyama, Toyonaka, Osaka 560-8531, Japan}

\author{M. Hagiwara}
 \affiliation{JST, Transformative Research-Project on Iron Pnictides, Tokyo 102-0075, Japan}
 \affiliation{KYOKUGEN, Osaka University, 1-3 Machikaneyama, Toyonaka, Osaka 560-8531, Japan}

\author{Y. Tomioka}
 \affiliation{National Institute of Advanced Industrial Science and Technology, Tsukuba 305-8568, Japan}
 \affiliation{JST, Transformative Research-Project on Iron Pnictides, Tokyo 102-0075, Japan}
 
\author{T. Ito}
 \affiliation{National Institute of Advanced Industrial Science and Technology, Tsukuba 305-8568, Japan}
 \affiliation{JST, Transformative Research-Project on Iron Pnictides, Tokyo 102-0075, Japan}
 
\author{S. Uchida}
 \affiliation{Department of Physics, University of Tokyo, Tokyo 113-0033, Japan}
 \affiliation{JST, Transformative Research-Project on Iron Pnictides, Tokyo 102-0075, Japan}

{\color{red}}

\begin{abstract}
\indent We investigated the transport properties of BaFe$_2$As$_2$ single crystals before and after annealing with BaAs powder. The annealing remarkably improves transport properties, in particular the magnitude of residual resistivity which decreases by a factor of more than 10. From the resistivity measurement on detwinned crystals we found that the anisotropy of the in-plane resistivity is remarkably diminished after annealing, indicative of dominant contributions to the charge transport from the carriers with isotropic and high mobility below magnetostructural transition temperature $T_{\rm s}$ and the absence of nematic state above $T_{\rm s}$. We found that the Hall resistivity shows strong non-linearity against magnetic field and the magnetoresistance becomes very large at low temperatures. These results give evidence for the manifestation of multiple carriers with distinct characters in the ordered phase below $T_{\rm s}$. By analyzing the magnetic field dependences, we found that at least three carriers equally contribute to the charge transport in the ordered phase, which is in good agreement with the results of recent quantum oscillation measurements. 
\end{abstract}

\maketitle

\section{\label{sec:level1}Introduction}

\indent Parent compounds of iron-based superconductors undergo a tetragonal-to-orthorhombic structural transition upon cooling accompanied by an antiferromagnetic (AFM) order and show anomalous metallic behavior.~\cite{Rotter} In this ordered phase, spins align antiferromagnetically along the $a$ axis and ferromagnetically along the $b$ axis,~\cite{Huang} so the electronic state is essentially anisotropic. Recent in-plane resistivity measurements performed on detwinned crystals of Ba(Fe$_{1-x}$Co$_x$)$_2$As$_2$, CaFe$_2$As$_2$, and SrFe$_2$As$_2$~\cite{Chu,Tanatar,Blomberg} revealed an anomalous anisotropic electronic properties that the resistivity along the AFM spin direction with longer lattice constant ($a$ axis) is smaller than the resistivity along the ferromagnetic (FM) spin direction with shorter lattice constant ($b$ axis) and this in-plane resistivity anisotropy can be seen in the tetragonal phase well above the phase transition temperature $T_{\rm s}$, which has been discussed in terms of nematicity.~\cite{Chu} In relation to the nematicity, the orbital polarization above $T_{\rm s}$ is observed by angle-resolved photoemission spectroscopy (ARPES).~\cite{Yi1} However, the in-plane resistivity above $T_{\rm s}$ is negligibly small in the cases of CaFe$_2$As$_2$ and SrFe$_2$As$_2$ which show a strongly first-order phase transition,~\cite{Blomberg} indicating that the in-plane resistivity anisotropy above $T_{\rm s}$ or nematicity might not be intrinsic but could be induced by the external pressure. \\
\indent The in-plane resistivity anisotropy of undoped BaFe$_2$As$_2$ is relatively small compared with underdoped Ba(Fe$_{1-x}$Co$_x$)$_2$As$_2$,~\cite{Chu} despite the lattice orthorhombicity diminishes monotonically with increasing Co concentration.~\cite{Prozorov} The small anisotropy of undoped BaFe$_2$As$_2$ has been discussed from magnetotransport measurements in terms of a dominant role of isotropic and high-mobility Dirac pockets with the Dirac point near the Fermi energy $E_F$,~\cite{Huynh,Kuo} which were first observed by ARPES (the Dirac point is located at 1~$\pm$~5~meV from $E_F$).~\cite{Richard} However, several Fermi surface (FS) pockets other than Dirac pockets also exist in the ordered state as evidenced by ARPES~\cite{Yi1,Yi2,Y.K.Kim} and quantum oscillation.~\cite{Terashima,Analytis} In such a multiple-band system, is it really possible that such tiny Dirac pockets with very small density dominate the charge transport? Additionally, recent optical measurement performed on detwinned BaFe$_2$As$_2$ crystal~\cite{Nakajima2} revealed that almost isotropic Drude component existing along both $a$ and $b$ axis dominates the dc conductivity. The estimated Drude weight is of the order of 10$^{20}$~cm$^{-3}$, which seems too large to be formed only by tiny Dirac pockets. It seems that one needs another explanation for the small in-plane resistivity anisotropy of undoped BaFe$_2$As$_2$. \\
\indent It is also reported that annealing remarkably improves transport properties in the ordered phase of BaFe$_2$As$_2$,~\cite{Rotundu} which indicates that the as-grown crystals contain appreciable amount of defects/impurities, and hence the observation of the intrinsic charge transport in this system might be inhibited. Therefore, it is required to investigate the intrinsic transport properties of this compound using annealed crystals. \\
\indent In this paper, we present the transport measurements on BaFe$_2$As$_2$ single crystals with much improved quality. We show that the in-plane resistivity anisotoropy both above and below $T_{\rm s}$ is diminished and the in-plane resistivity becomes almost isotropic, which indicates that the nematicity may not be intrinsic. We also provide clear evidence from the magnetotransport measurements that multiple carriers contribute to the charge transport in the ordered phase of BaFe$_2$As$_2$. Only tiny Dirac pockets do not play a dominant role as previously reported, but contribution to the conductivity from at least three types of carriers mask the underlying anisotropic electronic states. 

%\url{http://publish.aps.org/revtex4/}.

\section{\label{sec:level2}Experimental procedures}

\indent Single crystals of BaFe$_2$As$_2$ were grown by the self-flux method as described elsewhere.\cite{Nakajima1} The crystals were cut in a rectangular shape along the tetragonal [110] directions which become $a$ or $b$ axes in the orthorhombic phase. Typical crystal dimensions were 1.5~$\times$~1.5~$\times$~0.5~mm$^3$ with the shortest edge along the $c$ axis. Part of the crystals were sealed into an evacuated quartz tube together with Ba or BaAs powders and annealed for several days. A standard four-terminal method was used for the in-plane resistivity measurements on twinned crystals. The magnetoresistance (MR) and Hall resistivity $\rho_{xy}$ were measured with the electrical current along the $ab$ plane and the magnetic field along $c$ axis. For detwinning, the BaFe$_2$As$_2$ crystals were set into an uniaxial pressure cell and detwinned by applying compressive pressure along the tetragonal [110] direction.~\cite{Liang} The resistivity along the $a$ and $b$ axis were measured simultaneously using Montgomery method~\cite{Montgomery} without releasing pressure. The measurements in the magnetic fields up to 7~T were performed in a Quantum Design physical property measurement system (PPMS). MR and $\rho_{xy}$ in the magnetic fields up to 14~T were measured at the High Magnetic Field Laboratory, KYOKUGEN, Osaka University.

\section{\label{sec:level3}Results and discussion}

\subsection{Annealing effect on in-plane resistivity}

\begin{figure}[t]
\includegraphics[width=0.4\columnwidth,clip]{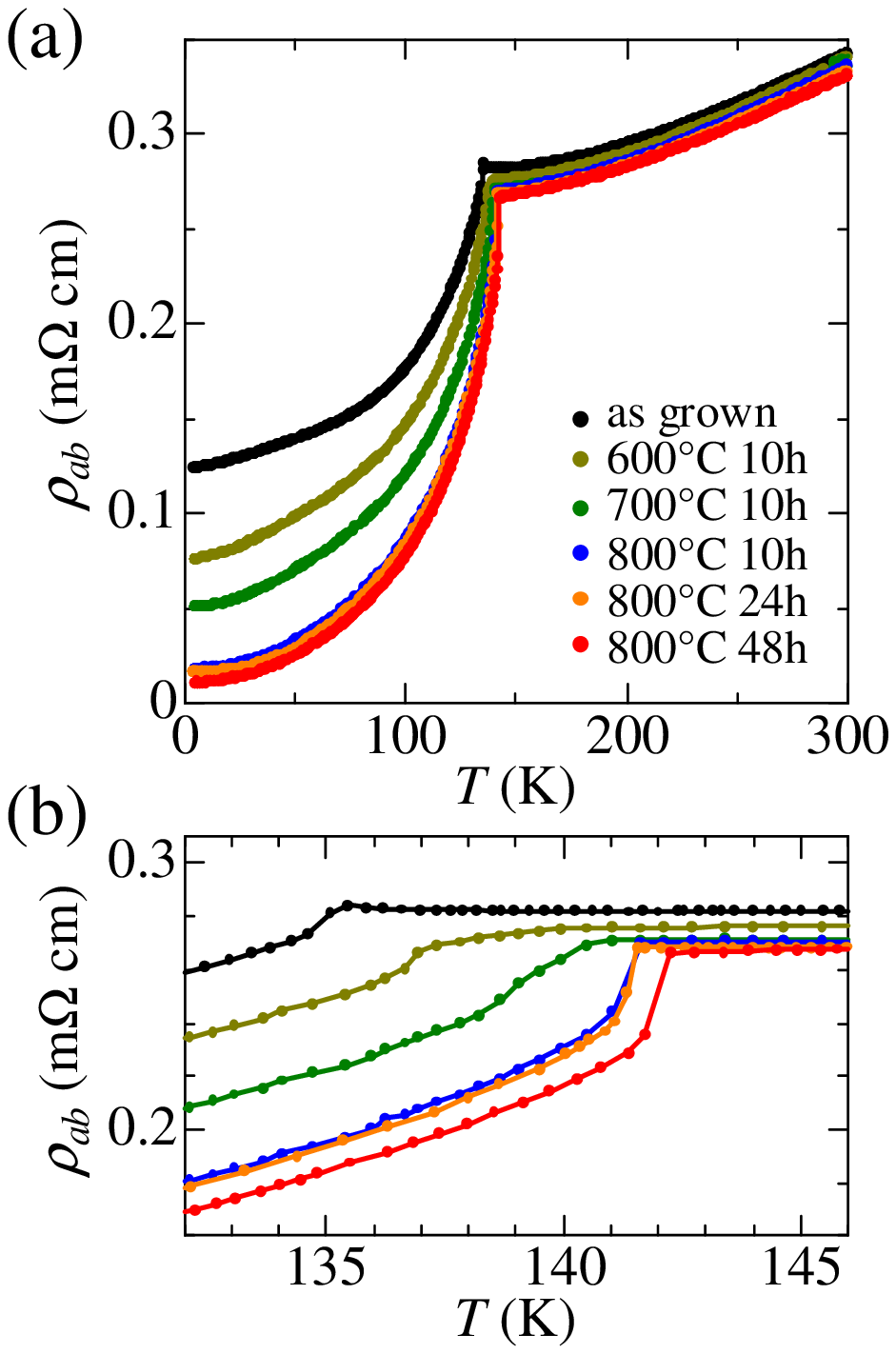}
\caption{\label{fig1} (Color online) (a) Temperature dependence of the in-plane resistivity measured on twinned BaFe$_2$As$_2$ crystals before (as grown) and after annealing under various conditions shown in the figure. Single crystals were annealed together with BaAs. (b) Enlarged view of (a) around the magnetostructural transition temperature.}
\end{figure}

\indent Figure \ref{fig1}(a) shows temperature ($T$) dependence of the in-plane resistivity $\rho_{ab}$($T$) measured on twinned BaFe$_2$As$_2$ crystals annealed under various conditions. $\rho_{ab}$($T$) of the as-grown crystal (black dots) shows typical temperature dependence so far reported for this compound. $\rho_{ab}$($T$) drops at $T_{\rm s}$ $\sim$ 136~K corresponding to the magnetostructural transition and shows residual resistivity of typically 0.1~m$\Omega$cm. The residual resistivity ratio ($RRR$) defined as $\rho_{ab}$(300K)/$\rho_{ab}$(5K) is about 3. Because the ordered state below $T_{\rm s}$ has orthorhombic crystal structure, the twinned crystal has domains with different orientations hence the scattering at domain boundaries could be a source of residual resistivity. However, this is unlikely as a main source of scattering in the present case since $\rho$($T$) of detwinned BaFe$_2$As$_2$ still shows large residual resistivity (see dashed curves in Fig. \ref{fig2}(a)). \\
\indent The annealing remarkably reduced the residual resistivity in the ordered phase of BaFe$_2$As$_2$ as reported in Ref.~\cite{Rotundu}. In general, the annealing process would remove crystal defects and lattice dislocations which are possible scattering sources and can cause large residual resistivity. Rotundu~$et~al.$ adopted low-pressure Ar gas annealing and obtained crystals with $RRR$ $\sim$~36 after 30-day annealing at 700~$^{\mathrm{o}}$C.~\cite{Rotundu} In order to shorten the annealing time, we annealed crystals in a evacuated quartz tube together with Ba or BaAs powders since we assumed that the main sources of scattering are defects and/or dislocations of Ba and/or As sites. However, annealing with Ba is found to not much improve the crystal quality, since Ba reacts with crystals above 600~$^{\mathrm{o}}$C and $RRR$ gets worse. We found that $RRR$ can be improved more quickly by annealing with BaAs than Ar gas annealing. After annealing for 48 hours at 800~$^{\mathrm{o}}$C, we obtained crystal with $RRR$ $\sim$~34. However, it is still unclear why $RRR$ of this compound is so sensitive to the annealing process. \\
\indent Temperature dependence of $\rho_{ab}$($T$) for the annealed crystal seems almost unchanged above $T_{\rm s}$ after annealing, $i.e.$ the annealing process scarcely affects the transport properties in the PM phase. On the other hand, as is shown in Fig. \ref{fig1}(b), $T_{\rm s}$ increases from 136~K to 142~K and the phase transition becomes sharper probably because crystal defects are removed and it helps to stabilize the orthorhombic lattice and magnetic ordering. Remarkably, $\rho_{ab}$($T$) drops much steeper below $T_{\rm s}$ and goes down to $\sim$ 10~$\mu\Omega$cm at $T$ = 5~K, one order of magnitude smaller than the value of as-grown crystal. If the carrier density is unchanged after annealing as reported,~\cite{Rotundu} this indicates that carrier mobility is greatly enhanced by annealing. 

\subsection{In-plane resistivity anisotropy}

\begin{figure}[t]
\includegraphics[width=0.4\columnwidth,clip]{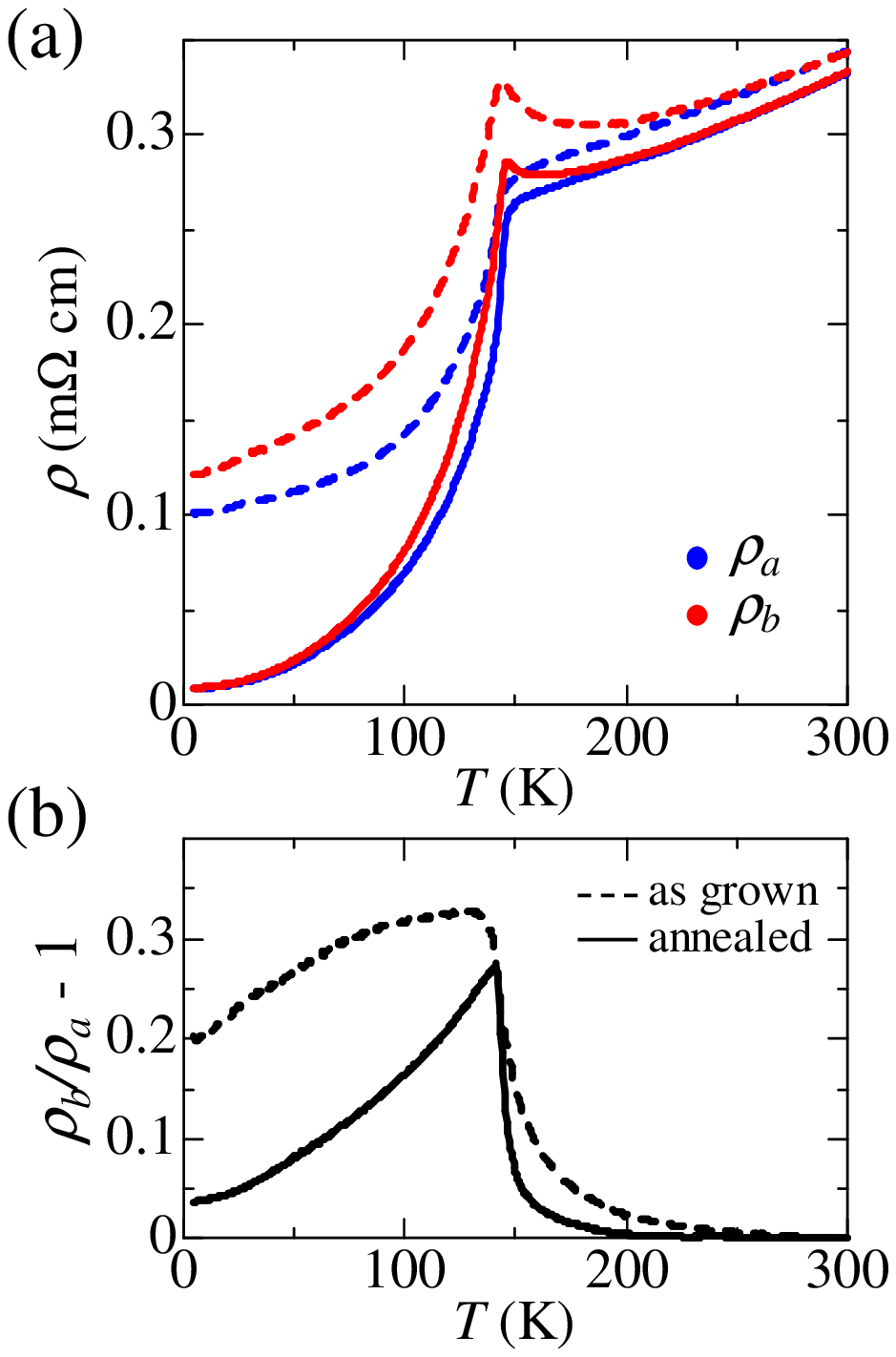}
\caption{\label{fig2} (Color online) (a) Temperature dependence of the in-plane resistivity measured on detwinned BaFe$_2$As$_2$ crystals, $\rho_{a}$ (blue) and $\rho_{b}$ (red). Dashed line and solid line correspond to as-grown and annealed crystal, respectively. (b) Temperature dependence of anisotropy of the in-plane resistivity defined as $\rho_b$/$\rho_a$-1 for as-grown (dashed line) and annealed (solid line) crystals.}
\end{figure}

\indent The in-plane resistivities of detwinned BaFe$_2$As$_2$ crystal are shown in Fig. \ref{fig2}(a). Note that effects of annealing on the in-plane resistivity anisotropy are also discussed by Nakajima,~$et ~al.$ in Ref.~\cite{Nakajima2}. The dashed line corresponds to resistivity components for the as-grown crystal. The blue and red colors correspond to resistivity along $a$ axis ($\rho_a$) and $b$ axis ($\rho_b$), respectively. Around room temperatures there is no appreciable difference between $\rho_a$ and $\rho_b$. As temperature is lowered, $\rho_a$ continues to decrease, while $\rho_b$ shows an upturn toward $T_{\rm s}$. The in-plane resistivity anisotropy can be seen well above $T_{\rm s}$, from about 80~K above, and $\rho_a$ is always smaller than $\rho_b$. These results reproduce the reported ones~\cite{Chu} and confirm that the crystal is successfully detwinned by our technique. The in-plane resistivity anisotropy defined as $\rho_b$/$\rho_a$-1 is plotted in Fig. \ref{fig2}(b). The anisotropy shows a maximum value $\sim$~0.33 just below $T_{\rm s}$, but it is robust even at the lowest temperature ($\sim$~0.2 at $T$ = 5~K). \\
\indent The solid lines in Fig. \ref{fig2}(a) show the resistivity components of the annealed crystal. The temperature dependence of $\rho_a$ and $\rho_b$ seems qualitatively similar to that of the as-grown crystal. However, the anisotropy becomes appreciable only about 40~K above $T_{\rm s}$, that is, the phase transition becomes more first-order like as seen in SrFe$_2$As$_2$ and CaFe$_2$As$_2$.~\cite{Krellner,Ni} This suggests that the magnetostructural transition might essentially be of the first order even in the purest sample of BaFe$_2$As$_2$. Therefore, the in-plane resistivity anisotropy above $T_{\rm s}$ seems not originated from nematic phase but induced by the external pressure and/or remaining crystal defects/disorder. Remarkably, the anisotropy is much reduces as compared with that for as-grown crystal and diminishes as temperature is decreased below $T_{\rm s}$ and the in-plane resistivity becomes nearly isotropic at temperatures well below $T_{\rm s}$ ($\rho_b$/$\rho_a$-1 $\sim$~0.04 at 5~K). In fact, the isotropic charge transport at low temperatures is consistent with the results of other experiments. ARPES~\cite{Yi1,Yi2,Y.K.Kim} and quantum oscillation~\cite{Terashima,Analytis} revealed ellipsoidal FS pockets which, when summed over the momentum space, would result in isotropic conductivity. Also, the optical conductivity spectra of detwinned BaFe$_2$As$_2$ for light polarization along $a$ and $b$ axis (measured crystals were annealed under the same condition)~\cite{Nakajima2} are dominated by an isotropic Drude component with extremely high peak value (of the order of 10$^5$~$\Omega^{-1}$cm$^{-1}$ at $T$ = 5~K) and narrow width, consistent with the dc resistivity. The anisotropy in optical conductivity shows up in finite frequency region, which arises from anisotropic gap feature below $T_{\rm s}$. The question is, whether or not this Drude term originates from the tiny Dirac pockets. We will answer this after we show the results of magnetotransport measurements. 

\subsection{Magnetoresistance}

\begin{figure}[t]
\includegraphics[width=0.8\columnwidth,clip]{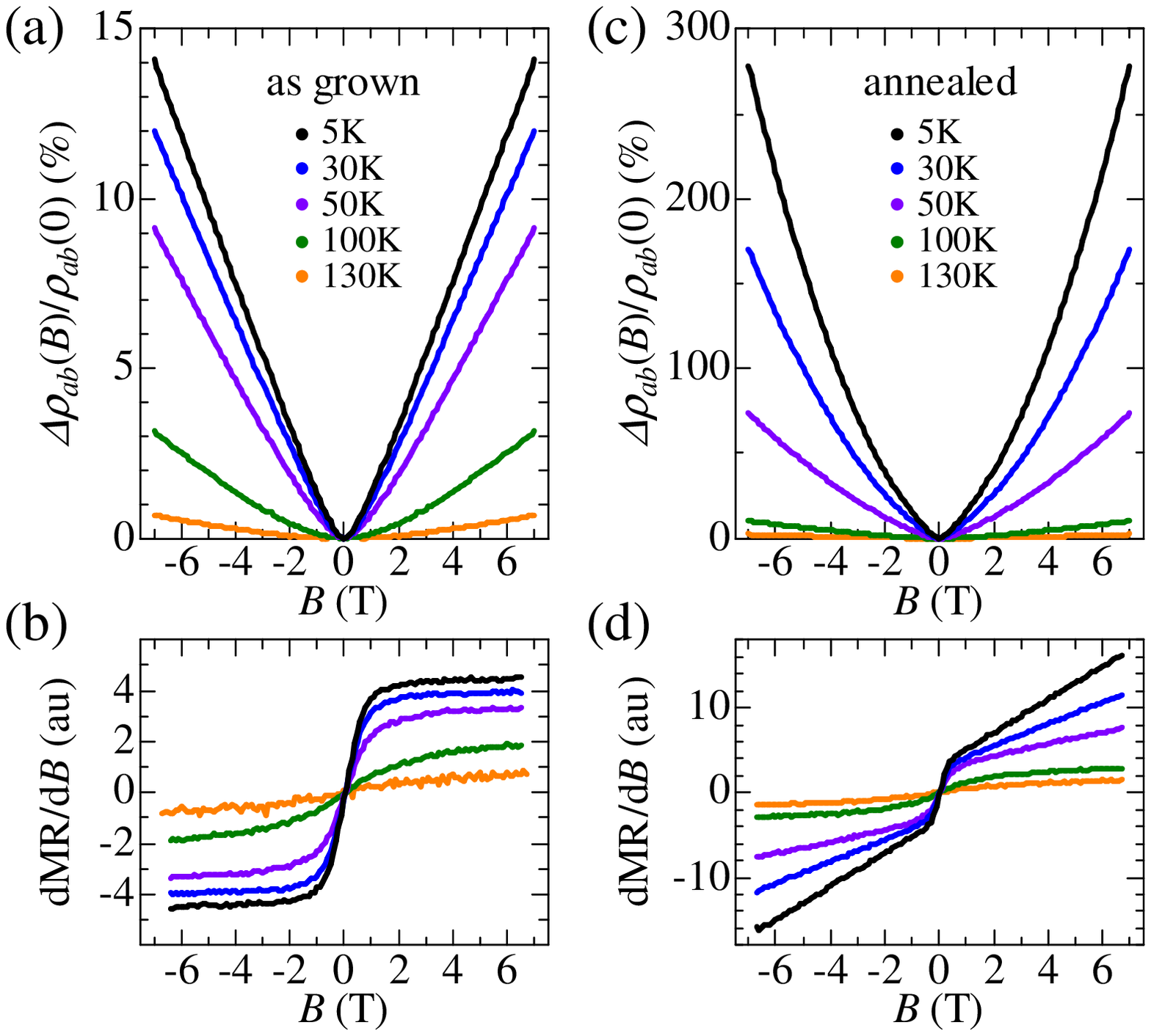}
\caption{\label{fig3} (Color online) Magnetic field dependence of the transverse magnetoresistnce (MR, $\Delta\rho_{ab}$($B$)/$\rho_{ab}$($B$=0)) (a) and its field derivative (dMR/d$B$) (b) for the twinned as-grown BaFe$_2$As$_2$ crystal taken at several temperatures. (c, d) Same plots for the annealed BaFe$_2$As$_2$ crystal.}
\end{figure}

\indent Figures \ref{fig3}(a)-(d) show the magnetic field ($B$) dependence of the magnetoresistance (MR) for transverse magnetic field ($B \parallel c$, $j \parallel ab$) and its derivative with respect to magnetic field for as-grown (a,b) and annealed (c,d) crystals, respectively. The magnitude of MR is defined as $\Delta\rho_{ab}$($B$)/$\rho_{ab}$($B$=0)=[$\rho_{ab}$($B$)-$\rho_{ab}$($B$=0)]/$\rho_{ab}$($B$=0). In both cases, MR is smaller than 0.03$\%$ in the PM phase and rapidly increases below $T_{\rm s}$. MR of as-grown crystal is 14~$\%$ at $B$ = 7~T and $T$ = 5~K (Fig. \ref{fig3}(a)) and shows almost $B$-linear dependence above 1T, which is more clearly seen in the $B$ dependence of the derivative of MR (dMR/d$B$) (Fig. \ref{fig3}(b)). $B$-linear MR is anomolous, since in the framework of simple two carrier model, MR at low fields can be written in the form of ($\mu_{\rm M}B$)$^2$ where $\mu_{\rm M}$ is magnetoresistance mobility, and MR saturates at high fields where $\mu_{\rm M}B$ or $\omega_c \tau$ $>$ 1. Huynh $et~al.$ ascribed this non-saturating linear MR to a quantum transport of Dirac cone states~\cite{Huynh} based on a model proposed by Abrikosov.~\cite{Abrikosov} However, recent study of MR for as-grown BaFe$_2$As$_2$ under higher magnetic field up to 50~T showed nonlinear but quadratic $B$ dependence of MR at high fields~\cite{Yuan} revealing the complicated $B$ dependence of MR. \\
\indent In the case of the annealed crystal, MR reaches $\sim$~280~$\%$ at $B$ = 7~T and $T$ = 5~K (Fig. \ref{fig3}(c)), which is by an order larger than that of as-grown crystal. Furthermore, in contrast to the case of as-grown crystal, MR shows quadratic $B$ dependence up to 7~T. Considering that the magnitude of MR is a measure of carrier mobility, the mobility of carriers should be greatly enhanced by the annealing, which is consistent with the decrease of residual resistivity. Note that dMR/d$B$ has a kink-like structure at low field around 0.5~T, that is, MR of annealed crystal also has some $B$-linear component similar to the case of as-grown crystal (Fig. \ref{fig3}(d)), suggestive of a contribution from Dirac pockets. However, observed large and quadratic $B$-dependent MR gives evidence that other FS pockets have appreciable contribution to the charge transport in the ordered phase of BaFe$_2$As$_2$.

\subsection{Hall effect}

\begin{figure}[t]
\includegraphics[width=0.8\columnwidth,clip]{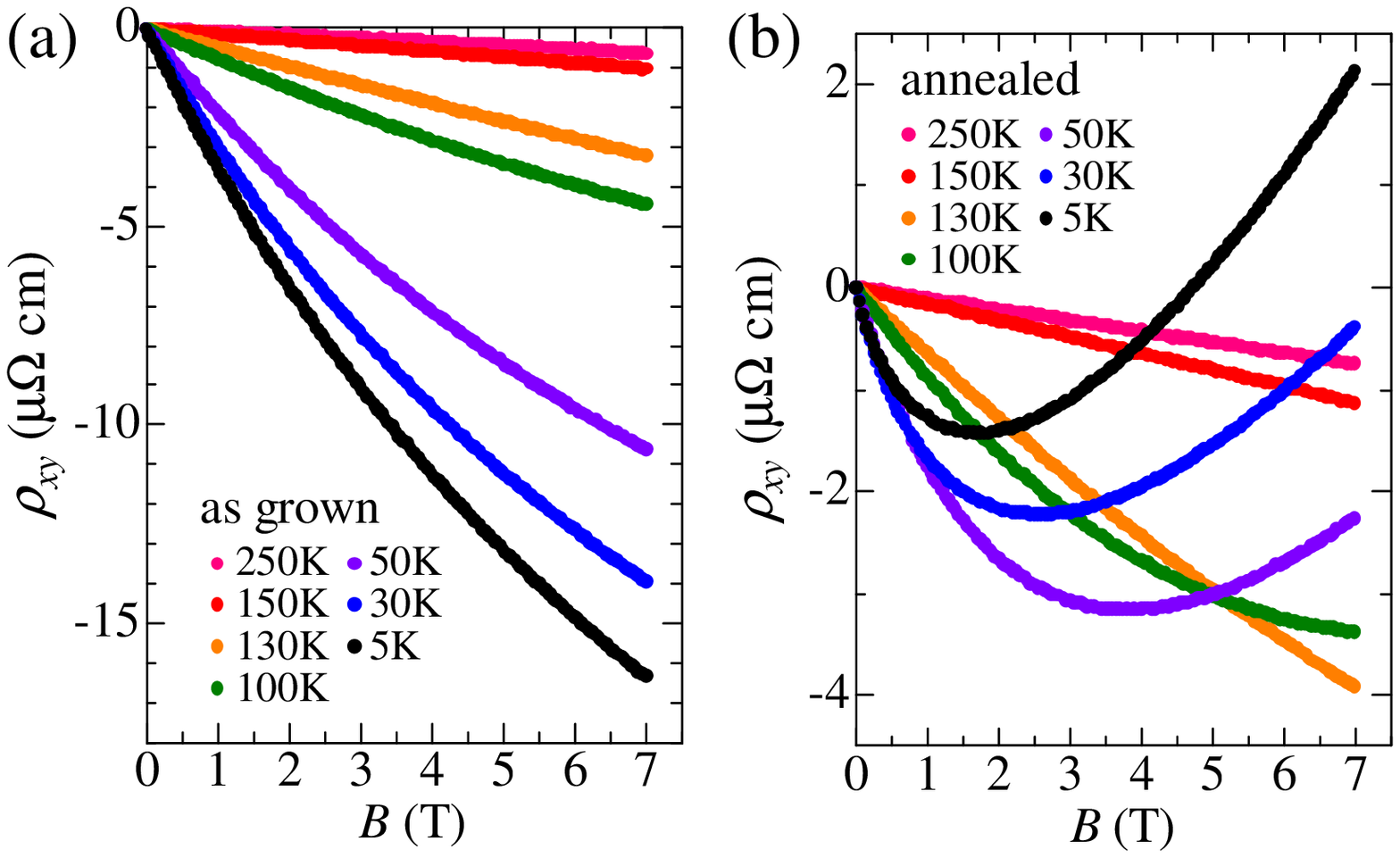}
\caption{\label{fig4} Magnetic field dependence of the Hall resistivity $\rho_{xy}$ of twinned as-grown (a) and annealed (b) BaFe$_2$As$_2$ crystals taken at several temperatures.}
\end{figure}

\indent Figure \ref{fig4}(a) shows the magnetic field dependence of the Hall resistivity $\rho_{xy}$ ($B \parallel c$, $j \parallel ab$) for as-grown crystal taken at several temperatures. $\rho_{xy}$($B$) of as-grown crystal shows basically $B$-linear dependence below $T_{\rm s}$ and a deviation from the $B$-linearity becomes apparent at $T$ well below $T_{\rm s}$. A previous report showed this sub-linear $B$ dependence of Hall resistivity persists to 50~T,~\cite{Yuan} which was ascribed to the $B$ dependence of carrier density. \\
\indent The $B$ dependence of $\rho_{xy}$($B$) for annealed crystal is shown in Fig. \ref{fig4}(b). In stark contrast to the as-grown crystal, $\rho_{xy}$($B$) below $T_{\rm s}$ soon deviates from $B$-linear dependence and exhibits nonmonotonic $B$ dependence. Especially at $T$ = 5~K, $\rho_{xy}$($B$) changes its sign to positive above $B$ $\sim$ 4~T. It is natural to attribute this $B$ dependence of $\rho_{xy}$($B$) to the multiple-carrier effect. Intuitively, the negative sign of Hall coefficient at low field should be related to electrons with higher mobility and the positive sign at high field indicates the prevalence of hole carriers with relatively lower mobility. \\
\indent Obviously, the magnitude of residual resistivity, the $B$ dependence of MR, and the nonmonotonic-$B$ dependence of $\rho_{xy}$ are all correlated. They orginate from multiple-carrier contribution which becomes clear as the sample quality is improved or as the carrier mobilities become higher. Below, in order to make more quantitative analysis on these results, we apply a multiple-carrier model to $\rho_{xy}$($B$). 

\subsection{Multiple-carrier model analysis}

\begin{figure}[t]
\includegraphics[width=0.7\columnwidth,clip]{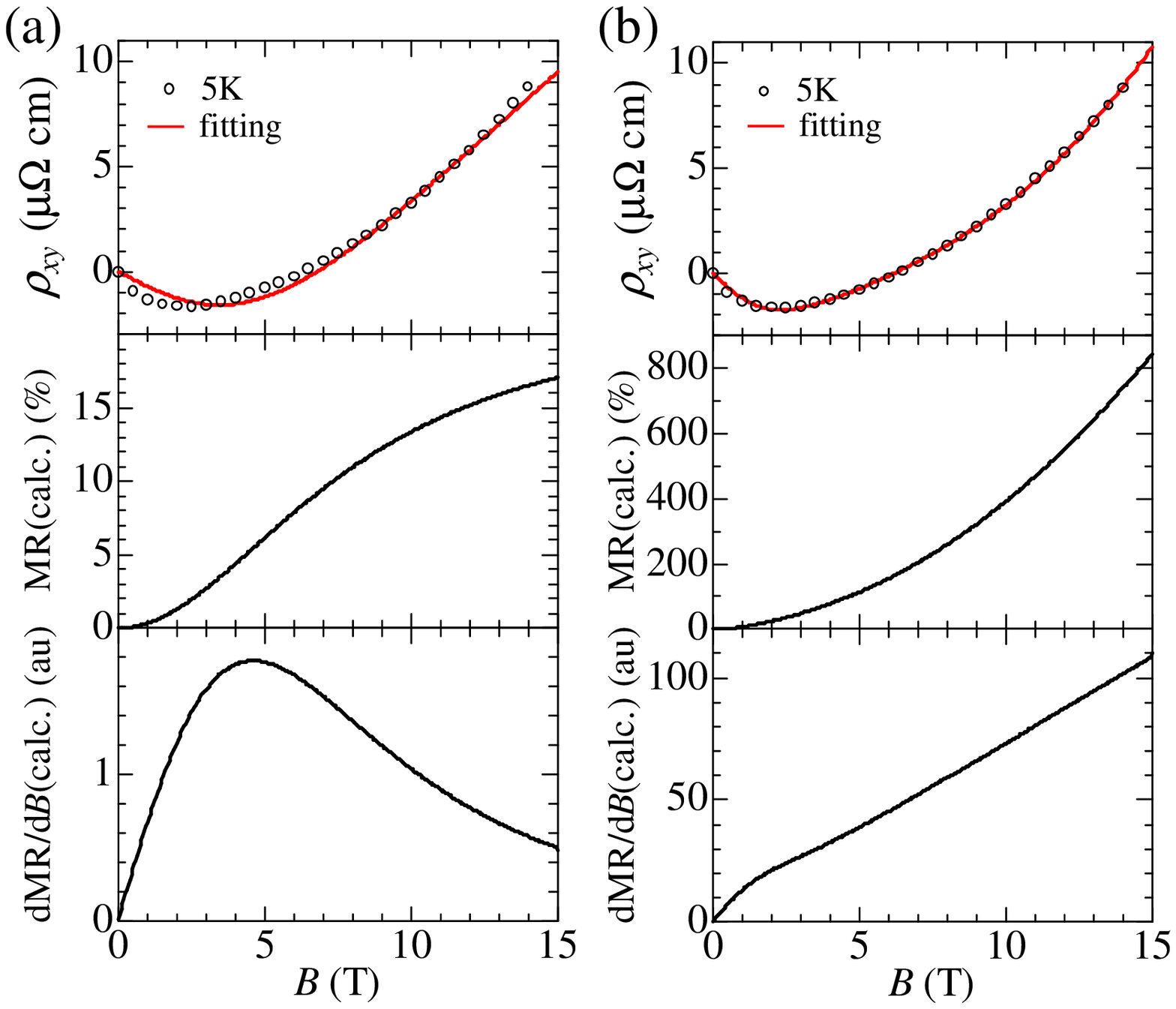}
\caption{\label{fig5} (Color online) (Top panel) Fitting results for $\rho_{xy}$($B$) of annealed BaFe$_2$As$_2$ crystal in the magnetic field up to 14~T at $T$ = 5~K using two-carrier (a) and three-carrier (b) model, respectively. (Middle panel) Calculated MR using the obtained parameter sets by fitting for $\rho_{xy}$($B$). (Bottom panel) Calculated dMR/d$B$.}
\end{figure}

\begin{figure}[t]
\includegraphics[width=0.4\columnwidth,clip]{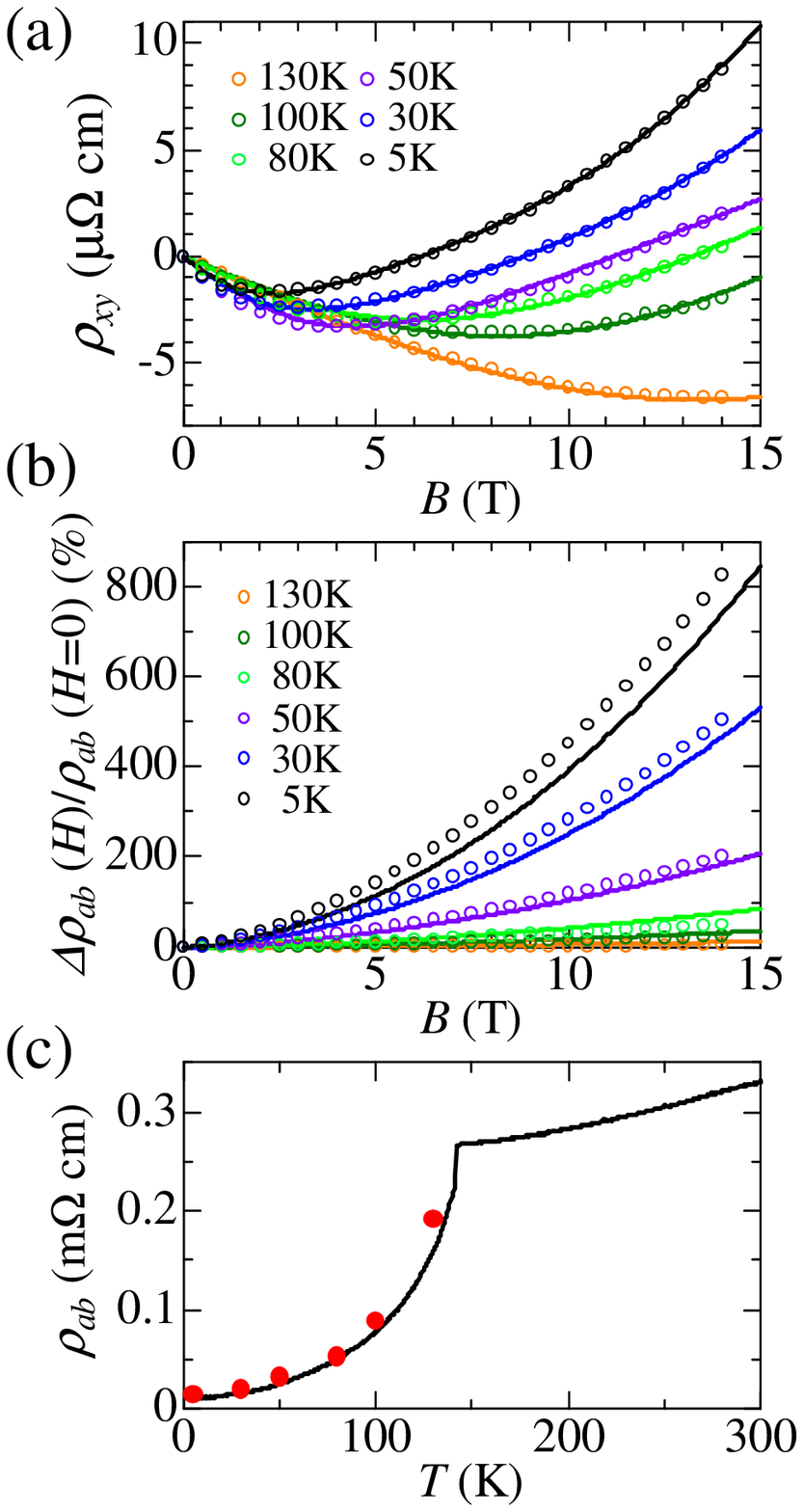}
\caption{\label{fig6} (Color online) (a) Fitting results for $\rho_{xy}$($B$) of annealed BaFe$_2$As$_2$ crystal using three-carrier model for various temperatures (solid lines). Open circles are experimental data. (b) Calculated MR using obtained parameter sets (solid lines). (c) Calculated in-plane resistivity (red dots). Black solid line is experimental data.}
\end{figure}

\indent First, we used a simple two-carrier model fitting~\cite{Chambers} assuming one electron-type and one hole-type carriers. As is shown in the top panel of Fig. \ref{fig5}(a), the two-carrier model fairly well fits $\rho_{xy}$($B$). However, the fitting parameter set shown in Table \ref{tab1} seems not reasonable, since (i) the density of holes is by two orders larger than that of electrons, inconsistent with ARPES~\cite{Yi1,Yi2,Y.K.Kim} and quantum oscillation~\cite{Analytis,Terashima}, (ii) a calculated MR using the obtained parameter set is $\sim$ 15~$\%$ at 14~T, much smaller than the experimental data ($\sim$ 800~$\%$ at $B$ = 14~T, see Fig. \ref{fig6}(b)), and (iii) the calculated MR shows saturating behavior which can be seen as decrease of dMR/d$B$ above 5~T (Fig. \ref{fig5}(c)). Any other parameter sets do not lead to better fitting than that shown in Fig. \ref{fig5}(a) (especially, if one put the constraint $n_e$~$\sim$~$n_h$, fitting of $\rho_{xy}$($B$) becomes worse). Thus, the two-carrier model is insufficient to describe the $B$ dependence of $\rho_{xy}$($B$) and MR. \\
\indent In order to explain the experimental data, we add a third carrier instead of considering $B$-dependent carrier density and/or Dirac cone states, namely, we apply three-carrier model fitting.~\cite{Chambers} (see Appendix A) One needs a number of parameters (carrier type, electron or hole, carrier density and mobility for each carrier and their anisotropy) for three-carrier model fitting, but there are several constraints; (i) anisotropy would be negligible since the in-plane charge transport is almost isotropic, (ii) total carrier density should be of order of 10$^{20}$~cm$^{-3}$ as estimated from the Drude weight in the optical spectrum,~\cite{Nakajima1} (iii) the densities of electrons and holes should be equal as is also confirmed by  the angle-resolved photoemission spectroscopy (ARPES)~\cite{Yi1,Yi2,Y.K.Kim} and quantum oscillation data~\cite{Terashima}, and (iv) the obtained parameter sets should be consistent with the magnitude of resistivity and MR. \\
\begin{table}[b]
\caption{\label{tab1} Values of parameters obtained by two- and three-carrier model fitting to the Hall resistivity of annealed BaFe$_2$As$_2$ at $T$ = 5~K in Figs. \ref{fig5}(a) and (b).}
\begin{ruledtabular}
\begin{tabular}{cccc}
carrier & 1 & 2 & 3 \\
\hline
type & $e$ (Dirac?) & $e$ & $h$\\
$n$ (2 carrier) (10$^{20}$cm$^{-3}$) & 0.05(4) & - & 6(2) \\
$\mu$ (2 carrier) (10$^3$cm$^2$/Vs) & 1.5(3) & - & 0.1(1) \\
$n$ (3 carrier) (10$^{20}$cm$^{-3}$) & 0.3(2) & 0.7(2) & 1.0(2) \\
$\mu$ (3 carrier) (10$^3$cm$^2$/Vs) & 4.5(5) & 1.5(2) & 1.8(2) \
\end{tabular}
\end{ruledtabular}
\end{table}
\indent In Fig. \ref{fig5}(b), we show the fitting results for $\rho_{xy}$($B$) at $T$ = 5~K. The obtained parameter sets are shown in the Table \ref{tab1}. From the result of fitting, we find a few important points. Firstly, $\rho_{xy}$($B$) of annealed crystal best fits when we assume one type of electron with very high mobility ($\mu_1^e$) and smallest density ($n_1^e$), another type of electron with relatively low mobility ($\mu_2^e$) and larger density ($n_2^e$), and holes with the largest density ($n_3^h$) and mobility comparable to the 2nd electrons ($\mu_3^h$). Secondly, it is possible to reproduce the temperature dependence of $\rho_{xy}$($B$) by assuming reasonable temperature dependence of mobility for each carrier without changing the carrier density (Fig. \ref{fig6}). Thirdly, $\rho_{xy}$($B$) of as-grown crystal best fits by suppressing the mobility of each carrier using the same carrier density obtained from the fitting for $\rho_{xy}$($B$) of annealed crystal (see Appendix B). As is shown in Table \ref{tab1} and Fig. \ref{fig7}, the three carriers have distinct characters each other but equally contribute to the charge transport. 

\subsection{Comparison with other experiments}

\indent One type of the electron carriers with highest mobility and smallest density we extracted from the present results is likely to be associated with electron FS pockets originating from Dirac-cone-like energy bands with the Dirac point at $\sim$ 23~meV below $E_F$ observed by ARPES measurements.~\cite{Y.K.Kim} Note that there are two types of Dirac-like pockets with different sizes in BaFe$_2$As$_2$ as observed by ARPES,~\cite{Y.K.Kim} and the Dirac-like pockets we consider here are larger ones, which are different from the tiny Dirac pockets whose dominant contribution to the charge transport has been discussed in previous reports.~\cite{Huynh,Kuo} As for the 2nd electron and the hole FS pockets we suppose, they are actually observed by ARPES.~\cite{Yi1,Yi2,Y.K.Kim} The three types of carriers obtained in this study are in good agreement with the recent quantum oscillation measurement performed on annealed BaFe$_2$As$_2$ crystal.~\cite{Terashima} \\
\indent Since all the three carriers have high mobility as is shown in Table \ref{tab1} (even the lowest one, $\mu_2^e$ = 1.5~$\times$~10$^3$~cm$^2$/Vs at $T$ = 5~K), probably they together form a narrow Drude component in the optical conductivity spectra,~\cite{Nakajima2} which is in contrast to the previous reports suggesting the dominant role of tiny Dirac pockets.~\cite{Huynh,Kuo} \\
\indent In fact, the $B$ dependence of MR, especially the linear $B$ dependence seen in as-grown BaFe$_2$As$_2$ case, cannot be explained even in the framework of three-carrier model using the same parameter sets obtained by analyzing $\rho_{xy}$. In order to explain this behavior, we probably need to include the fourth carrier, which might be electrons with very small density ($\sim$~10$^{17}$~cm$^{-3}$) and exclusively high mobility ($\sim$~10$^4$~cm$^2$/Vs) perhaps originating from tiny Dirac pockets. However, the contribution, if any, to the conductivity from such pockets has to be small in the annealed samples, since the estimated conductivity ($\sim$~10$^3$~$\Omega$cm$^2$) is much smaller than those of three types of carriers we obtained in this study (Fig. \ref{fig7}(b)).

\section{\label{sec:level4}Conclusions}

\indent We found that annealing BaFe$_2$As$_2$ crystals with BaAs is very efficient way to improve the quality of the crystals. Using those annealed crystals, we observed strongly non-linear $B$ dependence of Hall resistivity and huge quadratic $B$-dependent magnetoresistance at low temperatures in the magnetostructurally ordered state. In order to understand these transport properties, we applied three-carrier, two types of electrons and one type of holes, model analysis and successfully reproduced the data quantitatively. These three types of carriers are in excellent agreement with recent quantum oscillation measurements made on similarly high-quality crystals. All the three carriers are found to equally contribute to the conductivity. From the measurements on detwinned crystals we found that the anisotropy of in-plane resistivity and its temperature range above $T_{\rm s}$ diminish as the sample quality is improved by annealing. This suggests that the so far conceived nematic phase is not intrinsic but induced by external pressure, impurities or crystal disorder. Also, the in-plane resistivity is almost isotropic at temperatures well below $T_{\rm s}$, even though the electronic state in the ordered phase is essentially anisotropic as evidenced by the optical conductivity. In the ordered state, a radical reconstruction of the FS takes place with gap opening in some part of the FS. Remaining carriers / reconstructed FS - two electron FS and one hole FS pockets - in the ordered state have isotropic contribution to the charge transport and mask the anisotropic charge dynamics at higher energies. 

\section*{\label{sec:level6}Acknowledgments}

\indent We thank T. Terashima and N. Kurita for helpful discussions. SI and MN thank to the Japan Society for the Promotion of Science (JSPS) for the financial support. This work was supported by Transformative Research-Project on Iron Pnictides (TRIP) from the Japan Science and Technology Agency, and by the Japan-China-Korea A3 Foresight Program from JSPS, and a Grant-in-Aid of Scientific Research from the Ministry of Education, Culture, Sports, Science, and Technology in Japan.

\section*{\label{sec:level5}Appendix}

\subsection{Fitting procedure}

\indent The fitting procedure is shown below. According to the three-carrier model,~\cite{J.S.Kim} the Hall resistivity is given by
\begin{equation}
\rho_{xy}(B) = \rho_{xx}(0) B (a + bB^2 + cB^4)/(1 + dB^2 + eB^4) \nonumber \\
\end{equation}
where
\begin{eqnarray}
&&\rho_{xx}(0) = 1/\sum_i|e n_i \mu_i| \nonumber \\
&&a = f_1 \mu_1 + f_2 \mu_2 + f_3 \mu_3 \nonumber \\
&&b = f_1 \mu_1 (\mu_2^2 + \mu_3^2) + f_2 \mu_2 (\mu_3^2 + \mu_1^2) + f_3 \mu_3 (\mu_1^2 + \mu_2^2) \nonumber \\
&&c = (f_1 \mu_2 \mu_3 + f_2 \mu_3 \mu_1 + f_3 \mu_1 \mu_2) \mu_1 \mu_2 \mu_3 \nonumber \\
&&d = (f_1 \mu_2 + f_2 \mu_1)^2 + (f_2 \mu_3 + f_3 \mu_2)^2 + (f_3 \mu_1 + f_1 \mu_3)^2 \nonumber \\
&&\hspace{6mm} + 2(f_1 f_2 \mu_3^2 + f_2 f_3 \mu_1^2 + f_3 f_1 \mu_2^2)^2 \nonumber \\
&&e = (f_1 \mu_2 \mu_3 + f_2 \mu_3 \mu_1 + f_3 \mu_1 \mu_2)^2 \nonumber \\
&&f_i = |n_i \mu_i|/\sum_i|n_i \mu_i| \nonumber  .
\end{eqnarray}
Note that $n_i$ and $\mu_i$ are the carrier density and mobility of $i$-th carrier, respectively, and the sign of the carrier mobility is negative for electrons and positive for holes in the calculation.

\subsection{Three-carrier model analysis of Hall resistivity of as-grown BaFe$_2$As$_2$ crystal}
\indent In Fig. \ref{fig8}(a), we show the results of three-carrier model fitting for the Hall resistivity of as-grown BaFe$_2$As$_2$ crystal in the magnetic field up to 50~T.~\cite{Yuan} Figure \ref{fig8}(b) shows the calculated MR using obtained parameter sets. We successfully reproduce those experimental data using three-carrier model, without introducing the $B$ dependence of carrier density. The mobility of each carrier is strongly suppressed compared with annealed BaFe$_2$As$_2$ crystal, but the carriers have almost equal contributions to the conductivity like in the case of annealed crystal (Fig. \ref{fig8}(c)). 
\begin{table}[t]
\caption{\label{tab2} Values of parameters obtained by three-carrier model fitting to the Hall resistivity of as-grown BaFe$_2$As$_2$ at $T$ = 5~K in the magnetic field up to 50~T.~\cite{Yuan}}
\begin{ruledtabular}
\begin{tabular}{cccc}
carrier & 1 & 2 & 3 \\
\hline
type & $e$ (Dirac?) & $e$ & $h$\\
$n$ (10$^{20}$cm$^{-3}$) & 0.2(2) & 0.7(2) & 1.2(2) \\
$\mu$ (10$^3$cm$^2$/Vs) & 1.1(3) & 0.3(2) & 0.2(2) \
\end{tabular}
\end{ruledtabular}
\end{table}

\newpage

\section*{\label{sec:level7}Figures}

\begin{figure}[t]
\includegraphics[width=0.4\columnwidth,clip]{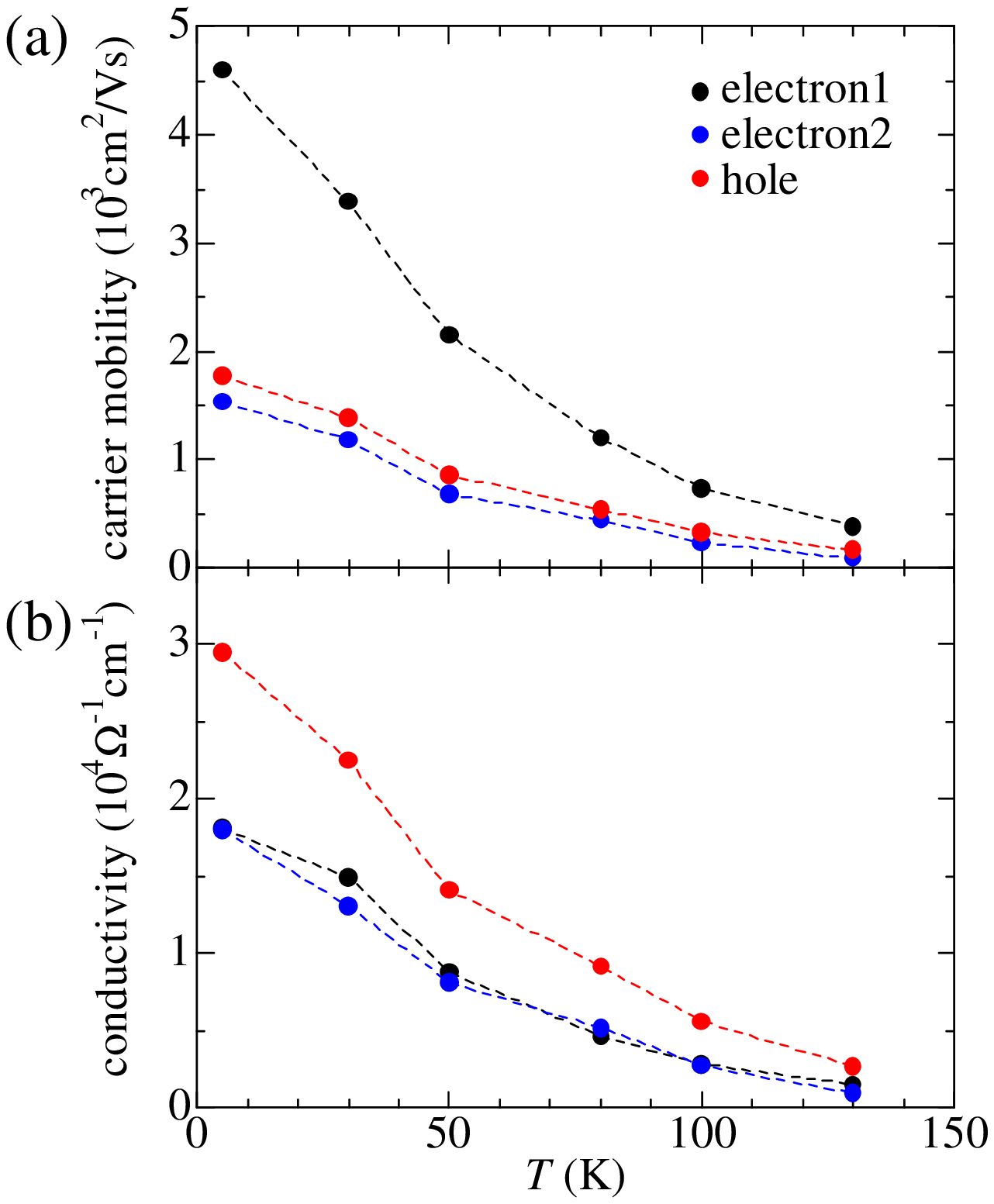}
\caption{\label{fig7} (Color online) Temperature dependence of carrier mobility (a) and conductivity (b) obtained by three-carrier model fitting.}
\end{figure}

\begin{figure}[t]
\includegraphics[width=0.4\columnwidth,clip]{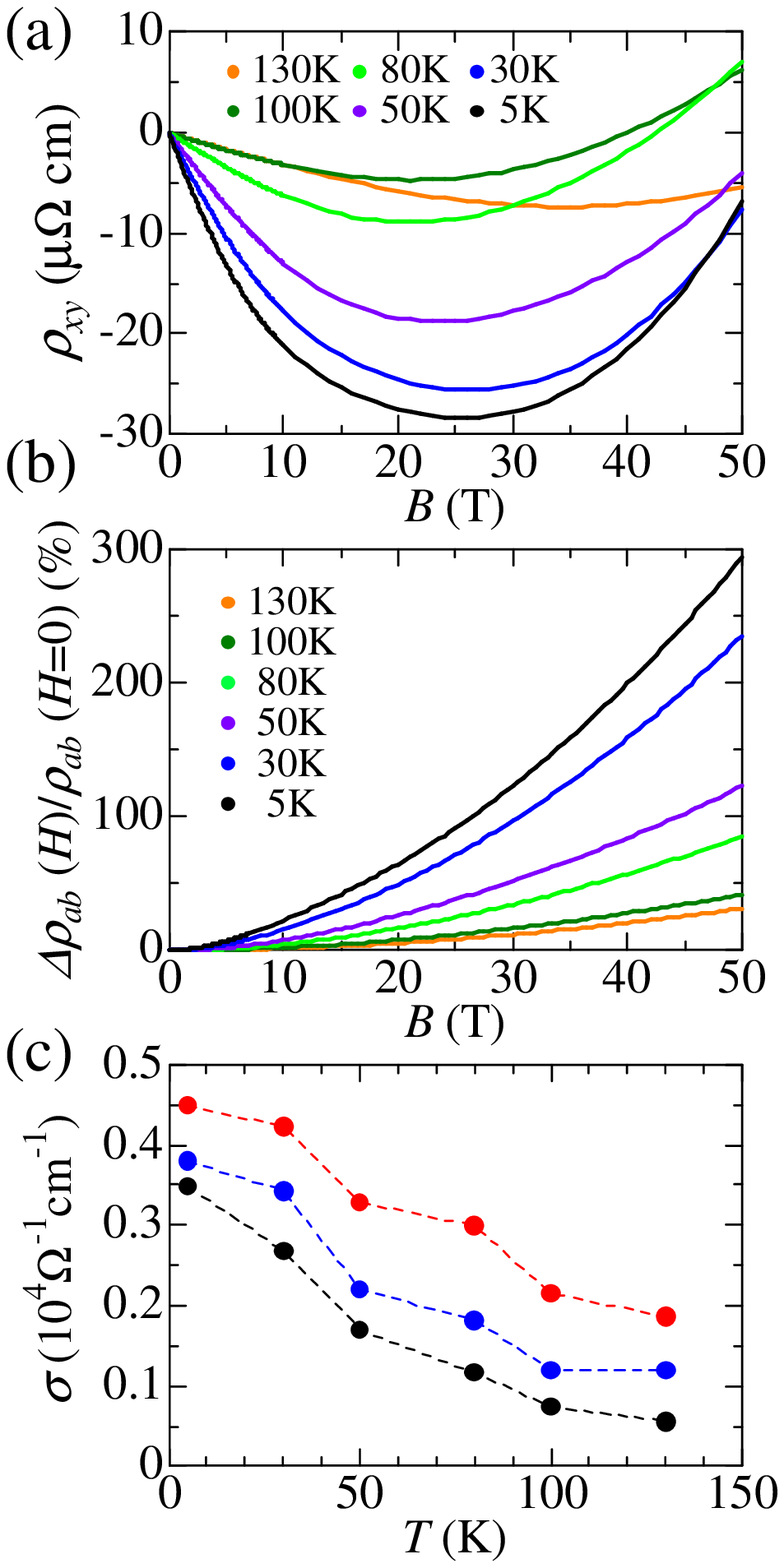}
\caption{\label{fig8} (Color online) (a) Fitting results for $\rho_{xy}$($B$) of as-grown BaFe$_2$As$_2$ crystal in the magnetic field up to 50~T~\cite{Yuan} using three-carrier model for various temperatures. Fitting parameters for $T$ = 5~K data are shown in Table \ref{tab2}. (b) Calculated MR. (c) Calculated conductivity $\sigma$.}
\end{figure}

\end{document}